# Superconductivity in CaBi$_2$

M.J. Winiarski[1,*], B. Wiendlocha[2], S. Gołąb[2], S. K. Kushwaha[3], P. Wiśniewski[4], D. Kaczorowski[4], J. D. Thompson[5], R. J. Cava[3], T. Klimczuk[1,†]

[1] *Faculty of Applied Physics and Mathematics, Gdansk University of Technology, Narutowicza 11/12, 80-233 Gdansk, Poland*

[2] *Faculty of Physics and Applied Computer Science, AGH University of Science and Technology, Aleja Mickiewicza 30, 30-059 Krakow, Poland*

[3] *Department of Chemistry, Princeton University, Princeton NJ 08544, USA*

[4] *Institute of Low Temperature and Structure Research, Polish Academy of Sciences, PNr 1410, 50-950 Wrocław, Poland*

[5] *Los Alamos National Laboratory, Los Alamos, New Mexico 87545, USA*

\* mwiniarski@mif.pg.gda.pl

† tomasz.klimczuk@pg.gda.pl

Superconductivity is observed with critical temperature $T_c$ = 2.0 K in self-flux-grown single crystals of CaBi$_2$. This material adopts the ZrSi$_2$ structure type with lattice parameters: $a$ = 4.696(1) Å, $b$ = 17.081(2) Å and $c$ = 4.611(1) Å. The crystals of CaBi$_2$ were studied by means of magnetic susceptibility, specific heat and electrical resistivity measurements. The heat capacity jump at $T_c$ is $\Delta C/\gamma T_c$ = 1.41, confirming bulk superconductivity; the Sommerfeld coefficient $\gamma$ = 4.1 mJ mol$^{-1}$ K$^{-2}$ and the Debye temperature $\Theta_D$ = 157 K. The electron-phonon coupling strength is $\lambda_{el-ph}$ = 0.59, and the thermodynamic critical field $H_c$ is low, between 111 and 124 Oe. Results of electronic structure calculations are reported and charge densities, electronic bands, densities of states and Fermi surfaces are discussed, focusing on the effects of spin-orbit coupling and electronic property anisotropy. We find a mixed quasi-2D + 3D character in the electronic structure, which reflects the layered crystal structure of the material.

*Keywords: Type-I superconductivity,*

# I Introduction

Much effort has recently been devoted to Bi-based candidates for topological insulators [1,2,3]. The presence of the heavy element bismuth provides the strong spin-orbit coupling that is essential for formation of the topologically-nontrivial band structure of these materials. Investigations also concentrate on finding candidates for a topological superconductor, with discussions currently primarily centering on $Cu_xBi_2Se_3$ [4,5]. There are several bismuth-alkaline and alkaline-earth metal intermetallic superconductors known, including LiBi [6], $NaBi$ [7], $KBi_2$ [8], $RbBi_2$, $CsBi_2$ [9], $Ca_{11}Bi_{10-x}$ [10], $SrBi_3$ [11,12], $BaBi_3$ [13], and $Ba_2Bi_3$ [14]. Recently, intermetallic phases in the Ca-Bi chemical system were examined using theoretical calculations, and three potential superconductors were proposed: $CaBi_3$, $CaBi$, and $Ca_3Bi$ [15]. The $CaBi_3$ phase, although found in many Ca-Bi phase diagrams [16,17,18], has not yet been described in detail. On the other hand, a $CaBi_2$ phase was reported by Merlo, et al. [19] but is not included in the Ca-Bi phase diagrams. A recent study of the calcium-bismuth system by means of electromotive force measurements and differential scanning calorimetry did not reveal the existence of either $CaBi_3$ or $CaBi_2$, although some traces of the latter were observed in scanning electron microscope analyses of arc-melted Ca-Bi alloys [20].

To the best of our knowledge, no report on the physical properties of $CaBi_2$ has been previously published. $CaBi_2$ crystallizes in an orthorhombic lattice in non-symmorphic space group *Cmcm* (no. 63) [19], and is isostructural with $ZrSi_2$ [21,22]. Fig. 1 shows its crystal structure, with Ca and Bi atoms marked by orange and violet balls, respectively. The structure has quasi-two-dimensional (2D) character, consisting of elongated, edge-shared Ca tetrahedra that are centered around Bi(2) atoms with Bi(1) atoms positioned almost in-plane with Ca layers perpendicular to the *b* axis. Alternatively, the structure can be viewed as a stacking of a square planar lattice of Bi(1) and a corrugated square lattice of Ca and Bi(2), with the nodes of one lattice positioned above the centers of the squares of the other lattice (see the right panel of Fig. 1). Such a quasi-2D crystal structure is expected to result in an anisotropy of electronic properties. The ZrSiS compound, in which the presence of Dirac cones was recently observed, has a structure that is closely related with those of $ZrSi_2$ and $CaBi_2$ [23, 24], and three $ZrSi_2$-type superconductors have been reported previously: $ScGe_2$ (1.3 K) [25], $LuGe_2$ (2.6 K) [26], and $YbSb_2$ (1.0 K) [27], the latter being the most extensively studied and claimed to exhibit type-I superconductivity [28, 29, 30].

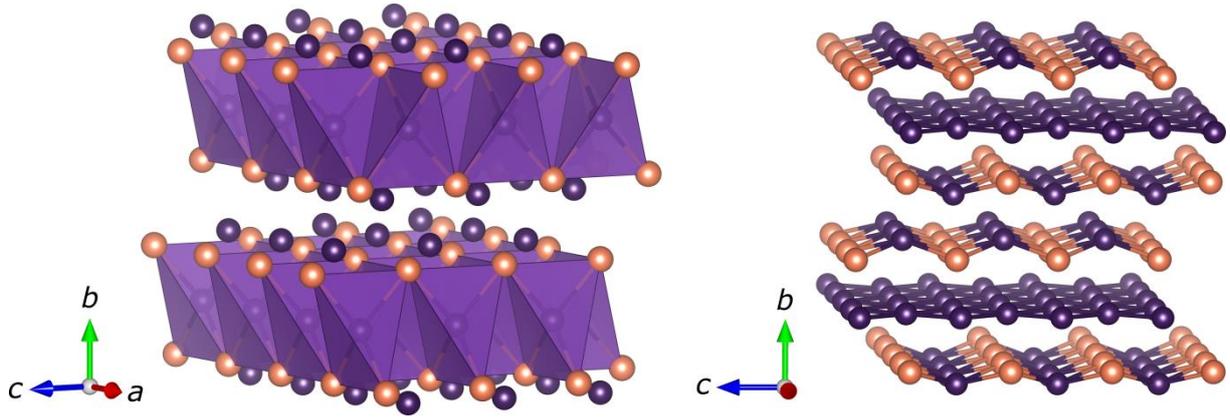

**Fig. 1** The CaBi$_2$ crystal structure. Calcium and bismuth atoms are represented by orange and violet balls, respectively. The Bi(2) atoms occupy the centers of Ca$_4$ tetrahedra that share long edges. The Bi(1) atoms are almost in plane with the Ca layer (left panel). Right panel an alternative view of the structure as a stacking of Bi(2) square planes and Bi(1)-Ca corrugated square nets (image rendered with VESTA [31]).

In order to study the electronic properties of CaBi$_2$, we employed a self-flux-based single crystal growth method [32], and the physical properties of the resulting crystals were analyzed. The electronic structure of the system was next calculated, using density functional theory methods. Electronic bands, densities of states, Fermi surfaces and charge densities are described here in addition to the electronic properties of the material; the analysis focuses on spin-orbit coupling effects and the anisotropy of the electronic states.

## II Materials and Methods

To grow the CaBi$_2$ crystals, calcium granules (Alfa Aesar, 99.5%) and bismuth pieces (Alfa Aesar, 99.99%) in a 3:17 molar ratio (15 at% of Ca) were put in a carbon-coated quartz tube inside an Ar-filled glovebox. A plug of quartz wool was then inserted, and the tube was subsequently evacuated and sealed without exposing the Ca metal to air. The ampule was heated to 550ºC, kept at that temperature for 8 hours, and then slowly cooled (3ºC per hour) to 310ºC at which temperature the excess Bi was spun off with the aid of a centrifuge (3000 rpm, relative centrifugal force ~ 1300). Clusters, of the diameter of the quartz tube used, composed of flat shiny crystals, were extracted. The lamellar morphology of the obtained crystals reflects the anisotropic crystallographic structure (see insets of Figs. 2 and 3). Single crystals were kept inside a glove box until characterized. Such a handling is necessary to avoid decomposition of CaBi$_2$.

The chemical composition of the crystals was examined with an FEI Quanta 250FEG Scanning Electron Microscope (SEM) equipped with an Apollo-X SDD Energy-Dispersive Spectrometer (EDS). EDS data were analyzed using the standardless analysis in the EDAX TEAM™ software. The phase identity of the crystals was examined using powder x-ray diffraction (PXRD) measurements employing a Bruker D8 FOCUS diffractometer with a Cu Kα radiation source and a LynxEye detector. A few single crystals were crushed for the PXRD measurements, with the powder covered with Paratone-N oil and a Kapton film to minimize sample degradation during the characterization. PXRD patterns were then analyzed by means of the LeBail [33] and Rietveld [34] methods using the FullProf software [35].

Heat capacity and electrical resistivity measurements were performed using a $^3$He-refrigerator equipped Quantum Design PPMS system. Electrical contacts were glued to the sample surface using silver paste. A standard relaxation method was used for the heat capacity measurements. Magnetic susceptibility measurements were carried out in a Quantum Design MPMS-XL SQUID magnetometer equipped with an iQuantum $^3$He refrigerator.

Electronic band structure calculations were performed using the plane-wave pseudopotential method and the QUANTUM ESPRESSO (QE) package [36]. Computations were done in the scalar-relativistic as well as relativistic (including spin-orbit coupling) modes. The Rappe-Rabe-Kaxiras-Joannopoulos ultrasoft pseudopotentials were used: scalar relativistic for Ca (electronic configuration $3s^23p^64s^2$) and for Bi (electronic configuration $5d^{10}6s^26p^3$), as well as relativistic for Bi (the same electronic configuration) [37]. The Perdew-Burke-Ernzerhof Generalized Gradient Approximation [38] (PBE-GGA) exchange-correlation (xc) potential was used. First, the unit cell dimensions and atomic positions in the reported orthorhombic *Cmcm* unit cell were optimized by minimizing the forces and stress, and next the charge densities, electronic band structures, densities of states and Fermi surfaces were computed and analyzed.

## III Results

The EDS results yielded the Ca:Bi ratio of 1:2, confirming the stoichiometry of the grown crystals. Additionally, some elemental Bi spots on the surface were found. They may originate either from remaining flux material that was not removed during the centrifugation process, or arise from $CaBi_2$ decomposition in contact with air and moisture. The room temperature PXRD pattern of crushed crystals is presented in Fig. 2 (the increased background in the low 2θ range is due to the Paratone-N oil). All the Bragg lines in the PXRD pattern can be indexed to orthorhombic *Cmcm* unit cell of CaBi$_2$ plus elemental Bi. (marked by green and red vertical bars, respectively). The Le Bail fit, represented by a black solid line in Fig. 2, gave for CaBi$_2$ the lattice parameters: $a$ = 4.696(1) Å, $b$ = 17.081(2) Å and $c$ = 4.611(1) Å. These values are in very good agreement with the data reported previously [19]. Details of the Le Bail analysis of the diffraction pattern, and lattice parameters from Ref. [19] are gathered in Table 1.The unit cell dimensions resulting from the energy minimization in the electronic structure calculations are slightly larger than those observed experimentally, as is often the case for GGA exchange-correlation potentials, resulting in about 2% bigger unit cell volume. The computed atomic positions (see Table 1) are very close to those in the isostructural ZrSi$_2$ compound [22]. An additional PXRD scan was performed on a sample exposed to air for 8 hours. The pattern showed that the CaBi$_2$ phase completely decomposed after that time period, with only Bi reflections observed.

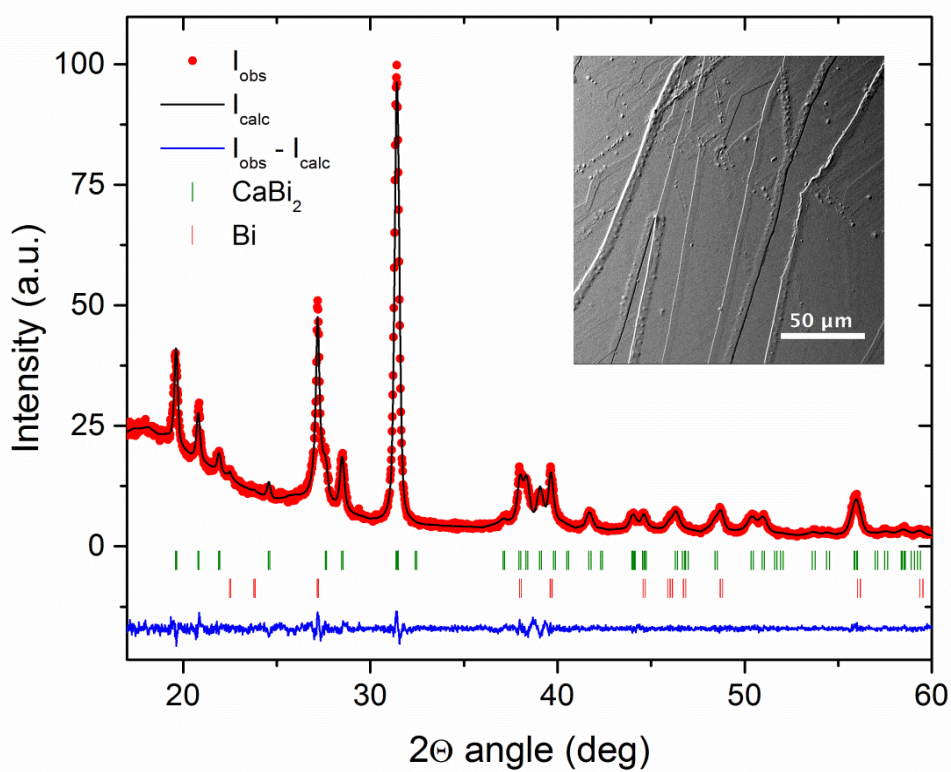

**Fig. 2 The X-ray powder diffraction characterization of crushed CaBi$_2$ single crystals. The plot shows the LeBail fit (black solid line) to the measured intensity (red dots). The blue line show the difference between measured and calculated intensity. Green ticks show the expected positions of reflections for the CaBi$_2$ phase, and the red tics show the positions for Bi. The inset presents a SEM photograph of a CaBi$_2$ single crystal surface showing a terrace-like topography. For the results of a Rietveld refinement on the same dataset see Fig. S1 and Table S1 of the Supplementary Information.**

The results of a XRD measurement on a freshly cleaved single crystal is shown in Fig. 3. Only reflections associated with (0 k 0), where k = even, were observed. This confirms that the surfaces of the lamellar crystals are perpendicular to the *b* crystallographic axis. The inset of Fig. 3 shows an optical photograph of a typical plate-like crystal.

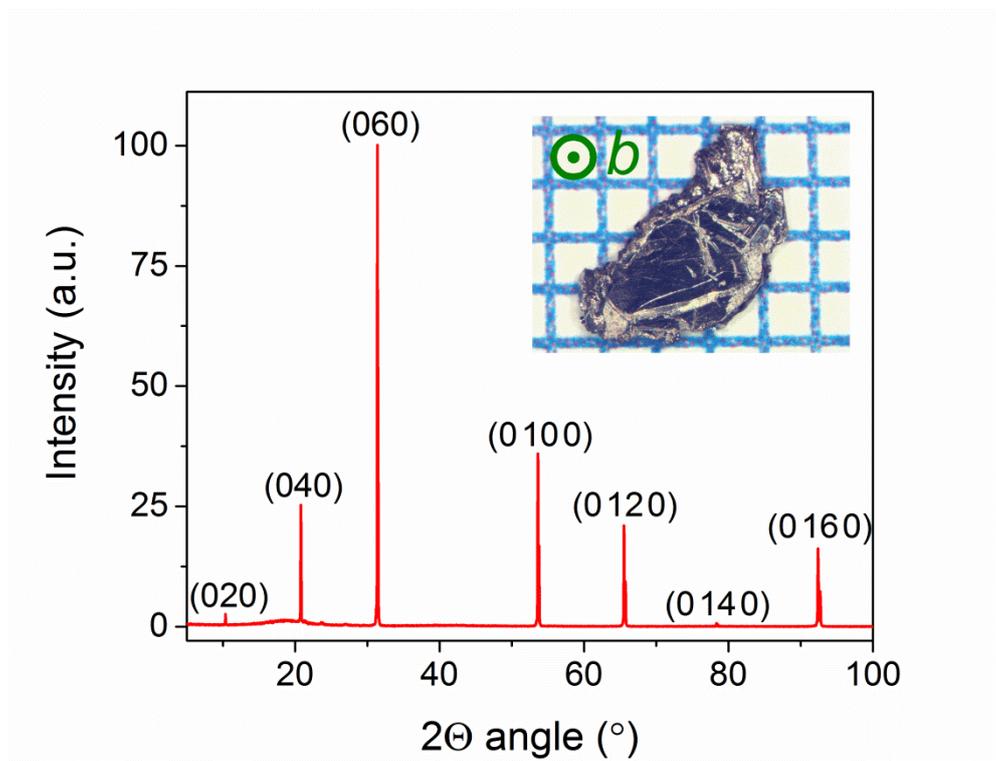

**Fig. 3 XRD pattern (Cu K$\alpha$ radiation, ambient temperature) collected on a single crystal plate. Only (0 k 0) reflections are observed, confirming that the orientation of the plate surface is perpendicular to the *b* direction. The inset shows the single crystal used for this XRD scan on a millimeter scale.**

**Table 1 Crystallographic data for CaBi$_2$.** Experimental data obtained from the LeBail fit to the measured XRD profile, and the computed parameters obtained by using the Quantum Expresso package with PBE-GGA xc-potentials. Cell parameters obtained from the LeBail fit are compared with values reported by Merlo and Fornasini [19]. For cell parameters and atomic positions resulting from a Rietveld fit to the diffraction data see Table S1 of the Supplementary Information.

| | CaBi$_2$ | | |
|---|---|---|---|
| Space group | *Cmcm* (# 63) | | |
| Pearson symbol | *oS*12 | | |
| Cell parameters (Å) | Le Bail fit: | Calculation: | Ref. [19]: |
| $a =$ | 4.696(1) | 4.782 | 4.701(2) |
| $b =$ | 17.081(2) | 17.169 | 17.053(6) |
| $c =$ | 4.611(1) | 4.606 | 4.613(2) |
| Atomic positions | | | |
| Ca | | (0, 0.0995, 0.25) | (0, 0.105, 0.25) |
| Bi(1) | | (0, 0.4332, 0.25) | (0, 0.44, 0.25) |
| Bi(2) | | (0, 0.7552, 0.25) | (0, 0.75, 0.25) |
| Cell volume (Å$^3$) | 369.9(2) | 378.2 | 369.8(4) |
| Molar weight (g/mol) | 458.04 | | |
| Number of formula units per cell – $Z$ | 4 | | |
| Density (calculated) (g/cm$^3$) | 8.225 | | |
| Figures of merit | | | |
| $R_p$ (%) | 17.2 | | |
| $R_{wp}$ (%) | 13.8 | | |
| $R_{exp}$ (%) | 11.1 | | |
| $\chi^2$ | 1.56 | | |

To characterize the superconducting transition of CaBi$_2$, zero-field cooling (ZFC) and field-cooling (FC) dc magnetization was measured in the temperature interval 0.48 K – 2.25 K. The plot of low-field ($H_{dc}$ = 5 Oe) volume susceptibility ($\chi_V$) vs. temperature is shown in Fig. 4(a). The superconducting transition temperature is $T_c$ = 1.95 K, defined as the temperature where the extrapolation of the steepest slope of $\chi(T)$ intersects the extrapolation of the normal state susceptibility to lower temperatures [39]. The superconducting transition is sharp and a full Meissner state is reached at ~1.7 K. The relatively small difference between the ZFC and FC curves suggests weak or nonexistent vortex pinning. Assuming a perfectly diamagnetic signal at the lowest temperatures for the ZFC data, one can estimate the value of the demagnetization factor

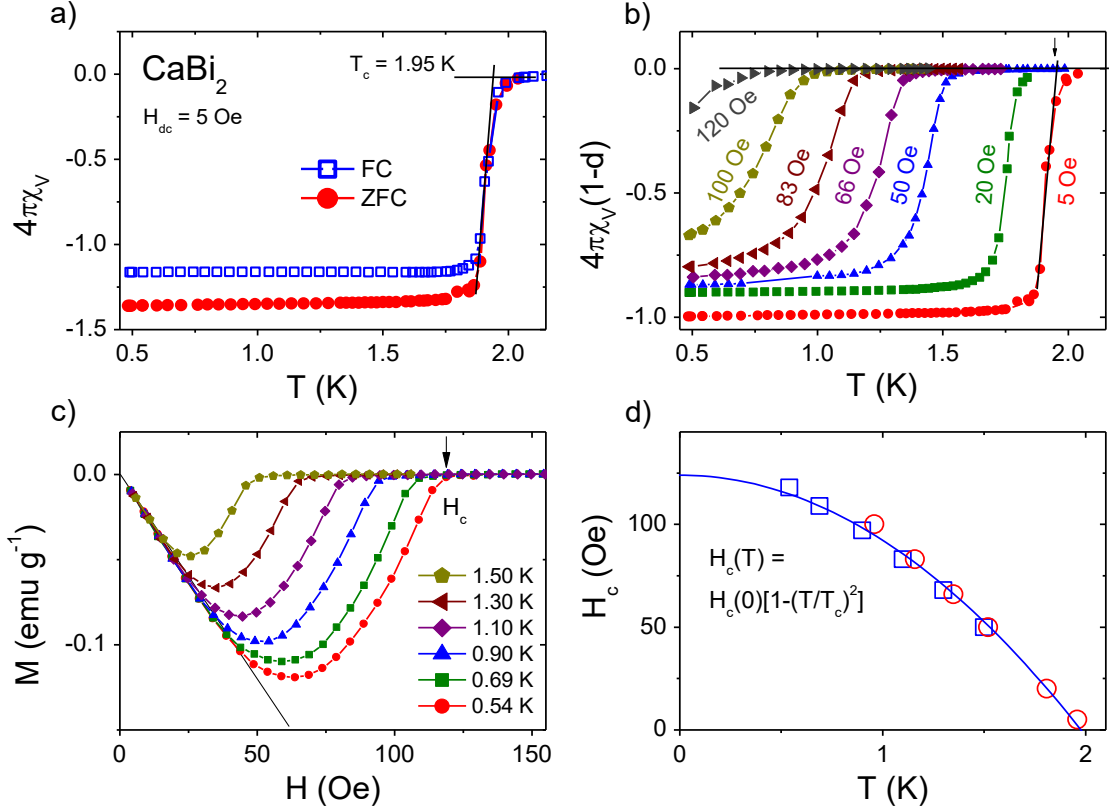

**Fig. 4 Magnetic characterization of the superconductivity in CaBi$_2$.** (a) ZFC (red circles) and FC (blue squares) volume magnetic susceptibility vs. temperature at a constant field of 5 Oe. (b) Volume magnetic susceptibility $\chi(T)(1-d)$ versus temperature in applied magnetic fields $H_{dc}$ = 5 Oe to 120 Oe. (c) Magnetization (*M*) versus applied magnetic field (*H*) at various temperatures. (d) Thermodynamic critical field $H_c$(T) determined from the $\chi(T)$ (circles) and *M*(*H*) (squares). The blue line is the $H_c(T) = H_c(0)[1-(T/T_c)^2]$ fit.

*d* using the relation: $-4\pi\chi_{V\,obs.} = \frac{1}{1-d}$, where $\chi_{V\,obs.}$ is the volume susceptibility at the lowest temperature. This calculations yield *d* = 0.27 and this number was used for plotting (panel c) temperature dependence of $\chi_V(1-d)$ vs. for different applied magnetic fields. With increasing field, $T_c$ decreases and the superconducting transition becomes broader. The obtained $T_c$ values were then used to estimate the thermodynamic critical field $H_c$(T). Figure 4(c) shows the magnetization (*M*) as a function of applied magnetic field (*H*) at various temperatures. The shape of the *M(H)* curves is almost symmetrical, and the high slope of *M(H)* close to the $H_c$ suggests type I superconductivity with influence of the demagnetization effect [40]. The variation of $H_c$ with temperature is shown in Fig. 4(d), where open circles are data points taken from the analysis of $\chi_V(T)$ and open squares are obtained from the $M(H)$ measurements. The whole range of the $H_c(T)$ data were fitted by using a formula $H_c(T) = H_c(0)[1 - \left(\frac{T}{T_c}\right)^2]$. As can be seen from Fig. 4(d), the fit (blue solid line) is very good and gives $H_c(0) = 124(2)$ Oe and $T_c = 1.98(2)$ K.

The specific heat measured through the superconducting transition in single-crystalline $CaBi_2$ is shown in Fig. 5(a). The bulk nature of the superconductivity is confirmed by a sharp, large anomaly at a temperature within error of those seen in susceptibility and resistivity. The specific heat jump at $T_c$, estimated by using the equal entropy construction method, is $\Delta C/T_c = 6.4$ mJ mol$^{-1}$ K$^{-2}$.

The panel (b) of Fig. 5 presents a plot of $C_p/T$ versus $T^2$, between 0.3 K and 2 K, measured in an applied magnetic field $H = 500$ Oe, that is higher than the thermodynamic critical field for $CaBi_2$ estimated from the magnetic susceptibility measurements. We fit the 0.9 K – 1.8 K region of $C_p$ to $C_p = \gamma T + \beta T^3$, where the first term is the normal state electronic contribution and the second one accounts for the lattice specific heat (in the Debye model). The black solid line is the resulting linear fit to the $C_p/T$ vs. $T^2$ data, which gives the Sommerfeld coefficient $\gamma = 4.6(1)$ mJ mol$^{-1}$ K$^{-2}$ and $\beta = 1.51(2)$ mJ mol$^{-1}$ K$^{-4}$. The Debye temperature is then calculated via the relation:

$$\Theta_D = \sqrt[3]{\frac{12\pi^4 nR}{5\beta}} \tag{3}$$

The resulting value of $\Theta_D$ is 157(1) K, which is about 40% larger than the Debye temperature of bismuth ($\Theta_D = 112$ K [41]). The sudden rise of $C_p/T$ below 0.7 K is likely caused by a nuclear Schottky contribution. The solid blue line represents data points measured in $H = 1$ kOe, clearly showing no difference from $C_p/T$ measured in 0.5 kOe.

Another important superconducting parameter is the energy gap ($2\Delta$) in the superconducting density of states, and it can be estimated by using a fully gapped (*s*-wave BCS) model $C_{el} \propto e^{-\Delta/kT}$. The electronic specific heat ($C_{el}$) reduced by a residual $\gamma_0 T$ term versus $T_c/T$ is presented in Figure 5(c). The solid orange line in panel (c) represents a fit to the data $\ln(C_{el}-\gamma_0 T) = a - \Delta/(k_B T)$. The small residual fit term ($\gamma_0 T$) of $0.5 \cdot T$ (mJ mol$^{-1}$ K$^{-1}$) originates from a small fraction of nonsuperconducting impurity phases and was held constant throughout the fit. An *s*-wave BCS model gives $\Delta_0 = 0.23(1)$ meV, which is close to $\Delta_0 = 0.3$ meV expected by the BCS theory $2\Delta_0 = 3.52\ k_B T_c$. The estimated energy gap for $CaBi_2$ is comparable to reported $\Delta_0 = 0.19$ meV for $YbSb_2$ [28], for which the occurrence of *s*-wave superconductivity is suggested by Nuclear Quadrupole Resonance (NQR) measurements [29].

By subtracting a residual $\gamma_0 = 0.5$ mJ mol$^{-1}$ K$^{-2}$ we find the Sommerfeld coefficient for $CaBi_2$ $\gamma_n = \gamma - \gamma_0 = 4.1$ mJ mol$^{-1}$ K$^{-2}$. Taking $\gamma_n = 4.1$ mJ mol$^{-1}$ K$^{-2}$, the normalized specific heat jump $\Delta C/\gamma T_c$ at $T_c$ can be estimated, yielding 1.41. This value is very close to the weak coupling superconductivity limit predicted by the BCS theory (1.43).

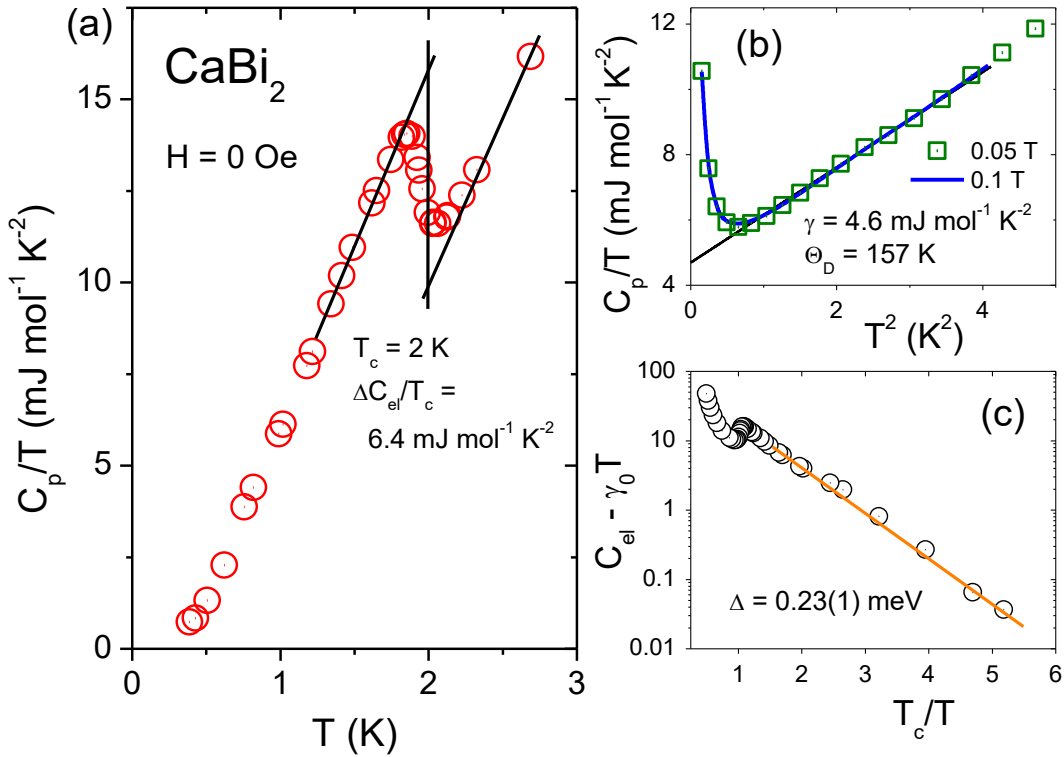

**Fig. 5. Low temperature specific heat of $CaBi_2$. (a) $C_p/T$ at 0 Oe magnetic field, exhibiting a clearly visible jump at $T_c$ = 2.0 K. The specific heat jump is estimated using the conserved entropy construction. (b) $C_p/T$ vs $T^2$ at 500 (red points) and 1000 Oe (blue line), above the upper critical field. A linear fit to the data is used to estimate the values of electronic and phonon specific heat coefficients. The rise of specific heat seen at the lowest temperatures, insensitive to magnetic field, can be attributed to the nuclear Schottky anomaly. Magnetic field was applied in the direction perpendicular to the *a-c* plane. (c) electronic specific heat ($C_{el} - \gamma_0 T$) versus $T_c/T$ with a $C_{el} \propto \Delta/(k_B T)$ fit. For specific heat results in wide temperature range (2-300 K) see Fig. S2 of Supplementary Information.**

Setting $\Theta_D = 157$ K and $T_c = 2.0$ K, and making the common assumption that the Coulomb pseudopotential parameter $\mu^* = 0.13$, we can now calculate the electron-phonon coupling constant $\lambda_{el-ph}$ using the modified McMillan formula [42]:

$$\lambda_{el-ph} = \frac{1.04 + \mu^* \ln\left(\frac{\Theta_D}{1.45 T_c}\right)}{(1 - 0.62\mu^*) \ln\left(\frac{\Theta_D}{1.45 T_c}\right) - 1.04} \quad (3)$$

This yields $\lambda_{el-ph} = 0.59$, indicating that $CaBi_2$ is a moderate-coupling strength superconductor. Having calculated the Sommerfeld coefficient and the electron-phonon coupling parameter the non-interacting density of states at the Fermi level $DOS(E_F)$ can be estimated using equation 4:

$$DOS(E_F) = \frac{3\gamma_n}{\pi^2 k_B^2 (1 + \lambda_{el-ph})} \quad (4)$$

The calculated $DOS(E_F) = 1.07$ states per eV per formula unit (f.u.).

The superconducting condensation energy ($\Delta U$) is given by $\Delta U = \frac{1}{2} DOS(E_F) \Delta_0^2$. Having the molar volume $V = 369.9$ Å$^3$, $Z = 4$, $DOS(E_F) = 1.07$ states / eV f.u. and $\Delta_0 = 0.23$ meV, we can estimate $\Delta U = 45.9$ J m$^{-3}$. The thermodynamic critical field ($H_c$) is related to the condensation energy by $\Delta U = \frac{1}{2} \mu_0 H_c^2$ and hence we estimate $H_c(0) = 111$ Oe which is in very good agreement with the thermodynamic critical field $H_c(0) = 124$ Oe obtained from the $H_c(T) = H_c(0)[1-(T/T_c)^2]$ fit (Figure 4(d)). This confirms our assumption that $CaBi_2$ is a type I superconductor.

The temperature dependence of the electrical resistivity for $CaBi_2$ is shown in Fig. 6. Due to the crystal morphology (thin platelets) the resistivity measurements could only be performed in the *a-c* plane. The normal state resistivity reveals metallic character with a high residual resistivity ratio $RRR = 200$. Such high values of the $RRR$ parameter have been reported before for single crystals of rare-earth dibismuthides grown by the Bi self-flux technique [32]. At the lowest temperatures, the measured resistance is very small and close to the resolution of the PPMS system. As shown in the inset of Figure 7, the midpoint of the superconducting transition is slightly above 2.0 K, and in the superconducting state the measured resistivity is finite. This may be attributed to a thin layer of elemental Bi on the surface that is a result of the partial decomposition of $CaBi_2$ during the preparation of the contacts for resistivity measurements.

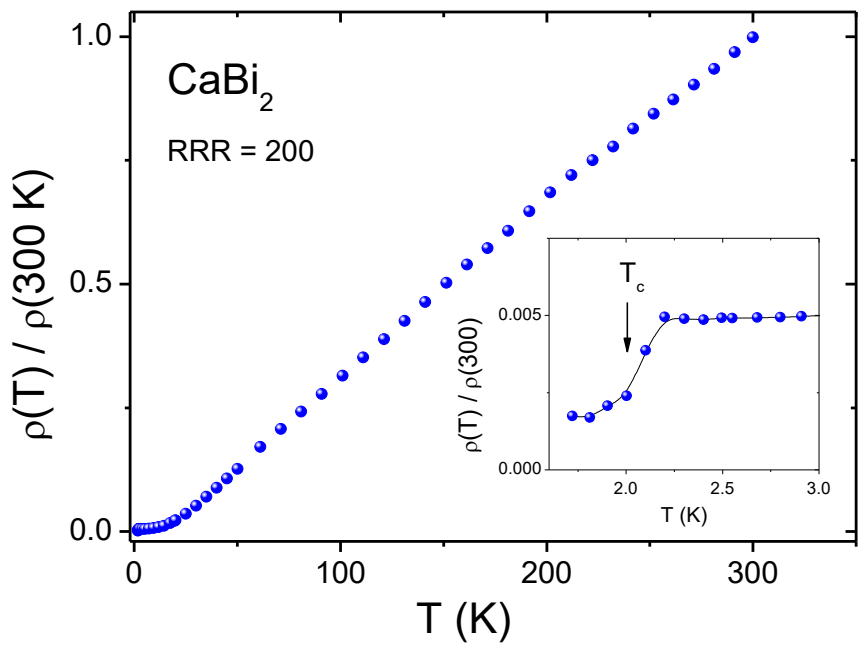

**Fig. 6. Electrical resistivity for CaBi$_2$. Main Panel, resistivity over a wide temperature range. Inset, the resistivity data near the superconducting transition in zero field ($H = 0$ Oe).**

**Electronic Structure Calculations**

We first analyze the charge density distribution corresponding to the electronic states included in the pseudopotentials. Fig. 7 presents these densities, computed including spin-orbit coupling, in various perspectives. Fig. 7(a) shows the primitive cell of CaBi$_2$ with the charge density projected on a "diagonal" plane, plotted in a logarithmic scale, in units e/a$_B^3$, ($e$ – electronic charge, a$_B$ – Bohr radius). Ca, as an alkaline earth metal, is a strongly electropositive element, thus one may expect that it transfers its 4$s$ valence electrons to bismuth. This is generally confirmed by the computations, since the electronic charge density between the Ca atoms is very small. Integration of the partial densities of states, discussed later, shows that one of the two 4$s$ electrons of Ca is transferred to the Bi atoms, approximately in an equal way, such that the resulting numbers of valence electrons in the main valence band are 1 for Ca and 3.5 for Bi(1) and Bi(2). This partial charge transfer shows the ionic nature of the bonding between Ca and Bi in CaBi$_2$. Visualization of the structure and charge density in Fig. 7(b) helps to distinguish the structural building blocks. The structure can be regarded as consisting of "sandwiches" of [Bi(1)-Ca]-[Bi(2)]-[Ca-Bi(1)], stacked along the y-axis ($b$ crystallographic axis if referring to the standard crystallographic unit cell). In Fig. 7(b) the charge density isosurface equal to 0.05 e/$a_B^3$ is rendered as a cloud surrounding the atoms. The "sandwiches" are connected mainly via Bi(1)-Bi(1) bonds. Figs. 7(c-j) show the densities and structures in the crystallographic unit cell. Panels (c-f) in Fig. 7 show the charge density projected on a set of planes, shown in (g-j), respectively. Again, a rather small electronic density is seen between Ca atoms, due to the large charge transfer from Ca to Bi (panels c and g). The largest electronic density is seen in the Bi(2) plane, located in the middle of the "sandwich" (panels e and i). In this "metallic" plane, the Bi(2) atoms form a centered square lattice. The Ca-Bi(1) planes (panels f and j), where Bi(1) atoms are slightly above/below the Ca plane (see the perspective view in panel b) also have the form of two square lattices, centering each other, and this plane has a more ionic character compared to that of the Bi(2). Finally, panels (d) and (h) shows the charge density in the middle between Bi(1) chains, which bound the "sandwiches". Analysis of the charge density shows that the system has a mixed 2D (layered) – 3D character from an electronic perspective, with the y-axis (crystallographic *b* axis) being special. Along this axis, we expect some impeded charge transfer.

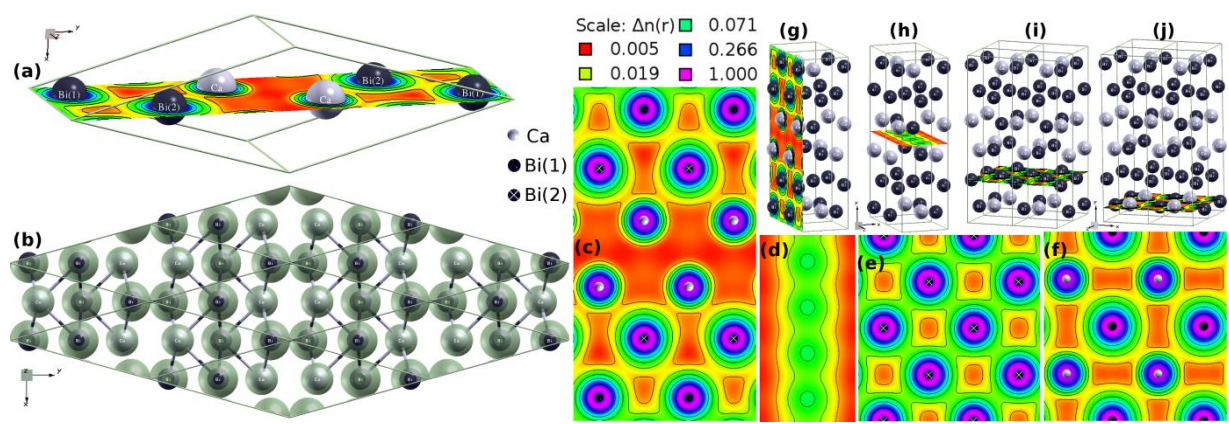

**Fig. 7. Charge density analysis of CaBi$_2$ (a) primitive cell of CaBi$_2$ with the calculated charge density projected on the [100] plane, plotted in logarithmic scale (units e/$a_B^3$, $a_B$ – Bohr radius); (b) perspective view of the structure, where "sandwiches" of [Ca-Bi(1)]-[Bi(2)]-[Ca-Bi(1)], stacked along y-axis, may be distinguished. The charge density isosurface, equal to 0.05 e/$a_B^3$, is rendered as a cloud surrounding the atoms. The "sandwiches" are connected mainly via Bi(1)-Bi(1) bonds (also projected in panel (h)). Panels (c-j) show densities and structures in a conventional crystallographic unit cell. Panels (c-f) show charge density projected on various planes, shown in (g-j), respectively. Analysis of the calculated charge density shows that the electronic system has a mixed 2D-3D character. Charge densities are plotted using XCrySDen [43].**

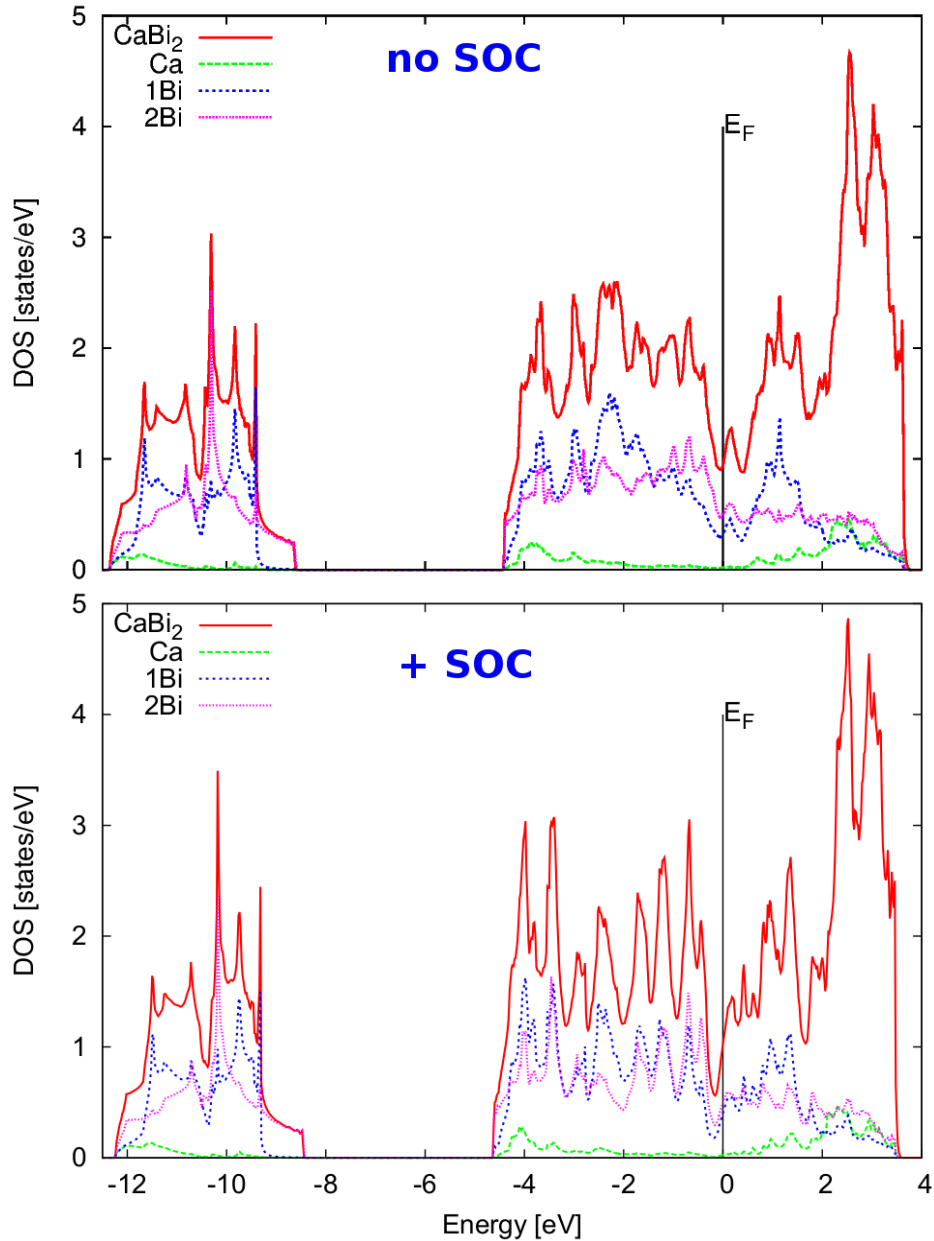

**Fig. 8. Calculated Electronic Density of States (DOS) of CaBi$_2$. Calculations performed without SOC (upper panel) and with SOC (lower panel). DOS is given per formula unit.**

Fig. 8 presents the calculated densities of states (DOS) in a broad energy range, given per formula unit, found in scalar-relativistic (upper panel) and relativistic (i.e. including spin-orbit coupling, lower panel) computations. CaBi$_2$ possesses 12 valence electrons per f.u. (two 4s electrons from Ca, two 6s and three 6p from each Bi). Since the *6s* orbitals of Bi are more strongly bounded to the core, the DOS associated with these states is separated and located deeper below E$_F$: between -12 eV and -9 eV, having only a small Ca contribution. That leaves 8 electrons in the main valence band (VB), i.e. between -5 eV and the Fermi level ($E_F$). The Ca atom, in agreement with its electropositive character and the charge density analysis, acts mainly as an electron reservoir, having its electronic DOS equally distributed over the valence band, and giving a small contribution to DOS near the Fermi level. Bismuth atoms play the key role in the electronic structure of the material, especially near $E_F$. As such, the strong influence of spin-orbit coupling (SOC) on the VB DOS is clearly seen on comparing the two panels in Fig. 8. The characteristic DOS peak near $E_F$, seen for the scalar-relativistic case is not equally pronounced when SOC is included, and a deep DOS valley below $E_F$ is formed. The value of the $DOS(E_F)$ changes slightly, from 1.07 (scalar-relativistic) to about 1.15 eV$^{-1}$ per f.u. (with SOC). The major contribution to DOS($E_F$) comes from the two Bi atoms, and thus these atoms are responsible for the superconductivity of the material. The computed DOS($E_F$) corresponds to the "bare" value of the Sommerfeld coefficient $\gamma_{calc}$ = 2.59 mJ mol$^{-1}$ K$^{-2}$ (see, Table 2), which, taking the experimental value $\gamma_{expt}$ = 4.1 mJ mol$^{-1}$ K$^{-2}$ allows for the estimation of the electron-phonon coupling constant $\lambda_{el-ph} = \frac{\gamma_{expt}}{\gamma_{calc}} - 1 = 0.58$. This value is in excellent agreement with $\lambda_{ep}$ = 0.59 deduced using the experimental T$_c$ and the McMillan formula. This confirms the classification of CaBi$_2$ as an intermediate coupling superconductor, however, with important spin-orbit coupling effects. The normal and superconducting parameters are listed in Table 2.

**Table 2 Values of the normal and superconducting parameters of CaBi$_2$. The $T_c$ is estimated from the specific heat measurement. Computed $DOS(E_F)$, $\gamma$, and $\lambda_{el-ph} = \frac{\gamma_{expt}}{\gamma_{calc}} - 1$ are also presented.**

| $T_c$ (K) | 2.0 | |
|---|---|---|
| $\gamma$ (mJ mol$^{-1}$ K$^{-2}$) | 4.1 (expt) | 2.59 (calc) |
| $\beta$ (mJ mol$^{-1}$ K$^{-4}$) | 1.51(2) | |
| $\Theta_D$ (K) | 157(1) | |
| $\Delta C_{el}/T_c$ (mJ mol$^{-1}$ K$^{-2}$) | 5.8 | |
| $\Delta C_{el}/\gamma T_c$ | 1.41 | |
| $H_c$ (Oe) | 111 – 124 | |
| $\lambda_{el-ph}$ | 0.59 (expt) | 0.58 (calc) |
| $DOS(E_F)$ (eV$^{-1}$ per formula unit) | 1.07 (expt) | 1.15 (calc) |

Fig. 9 shows electronic dispersion relations of CaBi$_2$, computed without SOC (left panel) and with SOC (right panel) in the *Cmcm* orthorhombic unit cell (containing 2 formula units). The four lowest bands are formed mainly by eight 6*s* electrons of bismuth, and the remaining 16 electrons are distributed over the next 10 bands, three of which cross the Fermi level. The strong spin-orbit coupling effects on the VB structure are well visible in all the presented directions, e.g. the second band crossing $E_F$ in the Y-T direction is considerably shifted by SOC, and the energies of some of the bands in Γ or Z points are shifted in about ~1 eV. Fig. 10 shows the band structure calculated with SOC around the Fermi level.

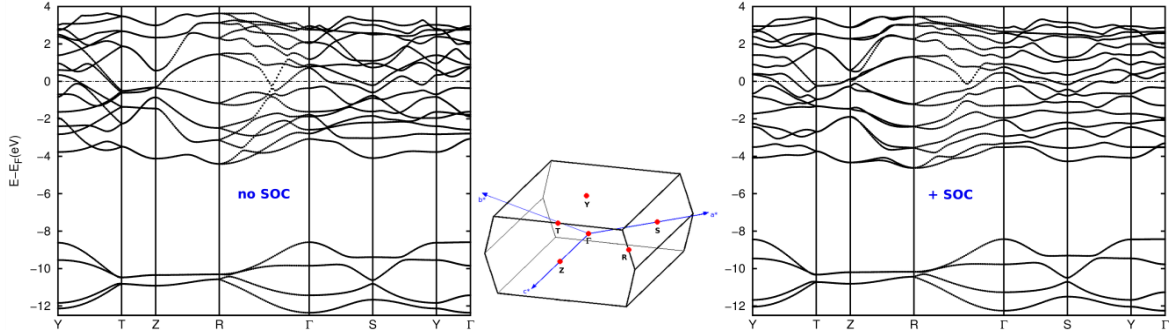

**Fig. 9. The calculated electronic band structure of $CaBi_2$, Computed without SOC (left panel) and with SOC (right panel). In the middle, the Brillouin zone (for the primitive cell with non-orthogonal $a^*$ and $b^*$ reciprocal base vectors), with the high symmetry points, is shown. The $k_y$ direction, which is parallel to the real-space b axis and perpendicular to the "sandwiches" is in the G-Y direction. $k_z$, which is parallel to c direction, goes along $c^*$ and $k_x$ is perpendicular to $k_y$ and $k_z$.**

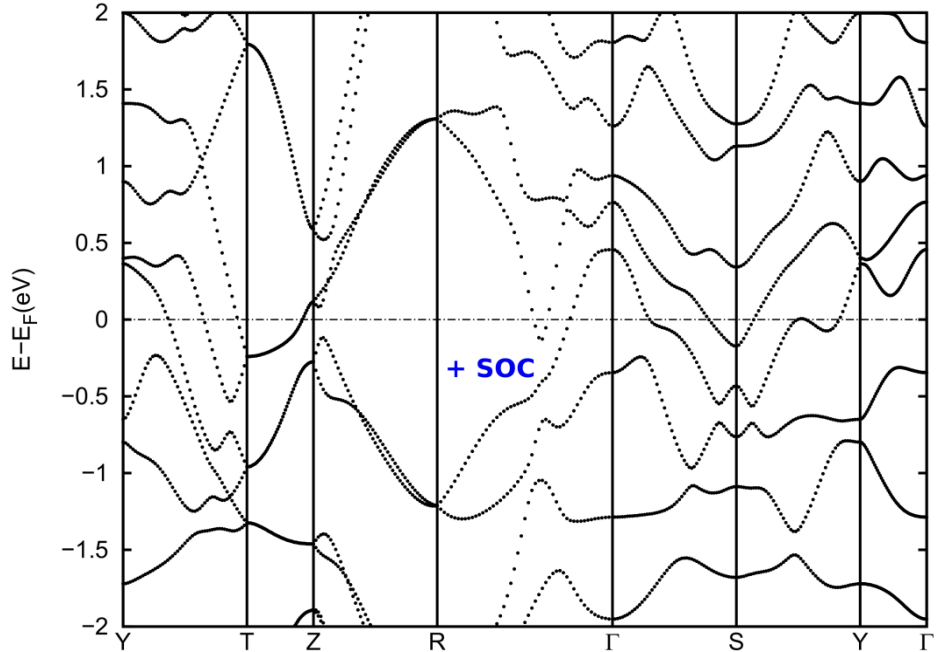

**Fig. 10. Relativistic band structure of $CaBi_2$ near the Fermi level. Brillouin zone and high symmetry points as shown in Fig. 9.**

The computed band structure reflects, to some extent, the geometry of the system, where layered (2D) and purely 3D elements are present. Since the computations are done in a primitive cell (not the crystallographic C-centered cell), the primitive $k$-space Brillouin zone, with non-orthogonal $a^*$ and $b^*$ vectors is shown in Fig. 9, and the high symmetry points are labeled using this convention. Nevertheless, the $k_y$ direction, which is parallel to the crystallographic $b$ axis, may be identified, and is represented by the Γ-Y direction, whereas the real space $a$-$c$ plane is perpendicular to Γ-Y. For such a layered structure we may expect the less-dispersive bands for the $k_y$ direction, which is perpendicular to the layers, and more dispersive bands in others. This is confirmed in Fig. 9, where the bands along T-Z and Γ-Y are much flatter than bands seen in other directions, especially in Y-T, which is parallel to the $a$-$c$ plane and the Bi-Ca-Bi sandwiches.

The calculated electronic band structure shows that CaBi$_2$ is an anisotropic, multi-band superconductor, and its Fermi surface consists of 3 sheets (3 bands are seen in Fig. 10, that cross $E_F$, e.g. in the Y-T direction). The computed Fermi surfaces are presented in Fig. 11. Panel (a) shows the scalar-relativistic FS, whereas panels (b)-(f) present the relativistic results. In panel (c), the FS is shown along the Γ-Y direction, which confirms its special character. Panels (d)-(f) show the FS sheets plotted separately.

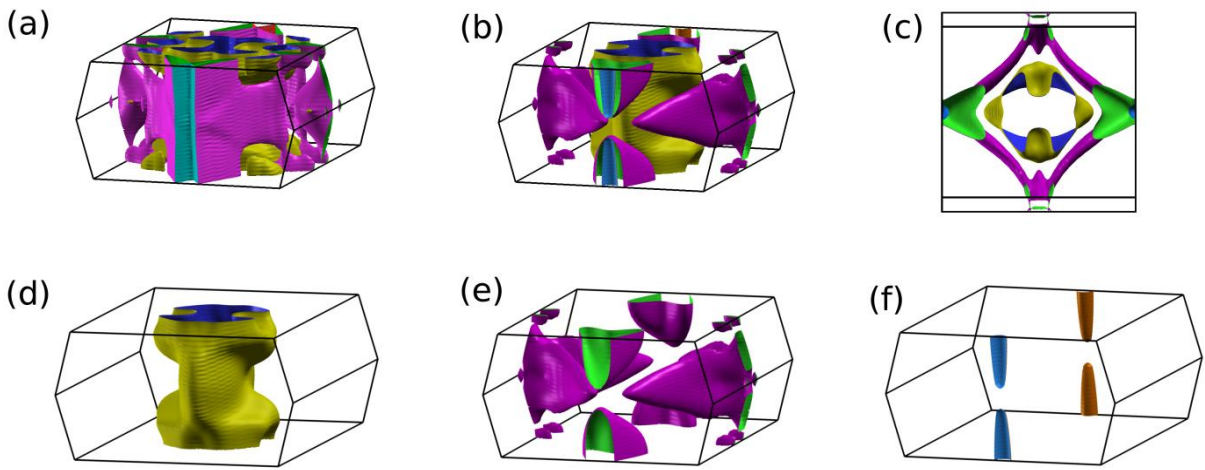

**Fig. 11. The Fermi surface of CaBi$_2$. (a) scalar-relativistic case, (b-f) relativistic case. Panel (c) shows a projection along Γ-Y direction, panels (d-f) show each of the three FS sheets: (d) quasi-2D electron-like cylinder, (e) eight 3D hole-like pockets and (f) four small hole-like ellipsoidal pockets. Bi(1) mainly contributes to the electron-like cylinder in (d), whereas Bi(2) to the hole-like pockets (e) and (f). Ca contribution is small and approximately equally distributed (see, also DOS pictures in Fig. 8). Pictures prepared**

using XCrySDen [43]. A comparison of the three sheets in scalar-relativistic and relativistic case is presented in Fig. S3 of Supplementary Information.

The Fermi surface of CaBi$_2$ is rather complex and consists of a one large, quasi-2D electron-like cylinder, centered along the $k_y$ line (Fig. 11(d), with relatively small dispersion along $k_y$), and two hole-like sheets: first with 8 large and complex-shaped pockets (Fig. 11(e)) and second with 4 ellipsoidal small pockets centered around the T point (Fig. 11(f)). Again, the strong influence of the spin-orbit coupling is seen, especially for the "second" (panel e) sheet, if compared to the appropriate sheet seen in scalar-relativistic panel (a). Bi(1) electronic states are mainly collected in the quasi-2D cylindrical sheet, whereas Bi(2) has larger contribution to the remaining parts of the FS. This mixed-dimensional character and multiple band structure of CaBi$_2$ may have an influence on its superconductivity.

## IV Conclusions

In conclusion, we have grown single crystals of CaBi$_2$ by a self-flux method. From magnetic susceptibility and heat capacity measurements, this material was found to be a phonon mediated superconductor with a transition temperature $T_c$ = 2.0 K. The field dependence of the magnetization in the superconducting state, plotted in Figure 4(c), suggests that CaBi$_2$ is a type I superconductor with significant influence of a demagnetization effect. The estimated thermodynamic critical field is low and its value is between 111 and 124 Oe.

Calculations showed that CaBi$_2$ is a three-band superconductor with strong spin-orbit interaction effects. Charge density computations showed that the crystal structure of CaBi$_2$ may be understood as consisting electronically of three-layer "sandwiches" of [Bi(1)-Ca]-[Bi(2)]-[Ca-Bi(1)] stacked along the $b$-axis connected via Bi(1) bridges. Such a structure naturally distinguishes the $b$-direction, which becomes the axis of symmetry for the quasi-2D electron-like Fermi surface pocket. The two hole-like FS pockets have more 3D-like character. The computed Sommerfeld parameter $\gamma$ allows us to estimate the electron-phonon coupling parameter $\lambda_{el-ph} = \frac{\gamma_{expt}}{\gamma_{calc}} - 1 = 0.58$, in close agreement with the value obtained from the measured critical temperature and McMillan formula, $\lambda_{el-ph} = 0.59$. This classifies CaBi$_2$ as a moderate strength electron-phonon coupling superconductor.

**Acknowledgements**

The research performed at the Gdansk University of Technology was supported by the National Science Centre (Poland) grant (DEC-2012/07/E/ST3/00584). B.W. was partially supported by the Polish Ministry of Science and Higher Education. The research at Princeton was supported by the Department of Energy Division of Basic Energy Sciences, Grant DE-FG02-98ER45706. Work at Los Alamos was performed under the auspices of the Department of Energy, Office of Basic Energy Sciences, Division of Materials Sciences and Engineering.

**References**

1 M. Z. Hasan and C. L. Kane, *Rev. Mod. Phys.*, 2010, **82**, 3045.
2 A. Isayeva, B. Raschke and M. Ruck, *Phys. Status Solidi RRL*, 2013, **1-2**, 39.
3 R. J. Cava, H. Ji, M. K. Fuccillo, Q. D. Gibson and Y. S. Hor, *J. Mater. Chem. C*, 2013, **1**, 3176.
4 Y. S. Hor, A. J. Williams, J. G. Checkelsky, P. Roushan, J. Seo, Q. Xu, H. W. Zandbergen, A. Yazdani, N. P. Ong, and R. J. Cava, *Phys. Rev. Lett.*, 2010, **104**, 057001.
5 N. Bray-Ali and S. Haas, *Physics*, 2010, **3**, 11.
6 T. Sambongi, *J. Phys. Soc. Jpn.* 1971, **30**, 294.
7 S. K. Kushwaha, J. W. Krizan, J. Xiong, T. Klimczuk, Q. D. Gibson, T. Liang, N. P. Ong and R. J. Cava, *J. Phys. Condens. Matter*, 2014, **26**, 212201.
8 S. Sun, K. Liu and H. Lei, *J. Phys. Condens. Matter*, 2016, **28**, 085701.
9 B. W. Roberts, *J. Phys. Chem. Ref. Data*, 1976, **5**, 581.
10 M. Sturza, F. Han, C. D. Malliakas, D. Y. Chung, H. Claus and M. G. Kanatzidis, *Phys. Rev. B*, 2014, **89**, 054512.
11 A. Iyo, Y. Yanagi, T. Kinjo, T. Nishio, I. Hase, T. Yanagisawa, S. Ishida, H. Kito, N. Takeshita, K. Oka, Y. Yoshida, H. Eisaki, *Sci. Rep.*, 2015, **5**, 10089.
12 D. F. Shao, X. Luo, W. J. Luo, L. Hu, X. D. Zhu, W. H. Song, X. B. Zhu and Y. P. Sun, *Sci. Rep.*, 2016, **6**, 21484.
13 N. Haldolaarachchige, S. K. Kushwaha, Q. Gibson and R. J. Cava, *Supercon. Sci. Technol.*, 2014, **27**, 105001.
14 A. Iyo, Y. Yanagi, S. Ishida, K. Oka, Y. Yoshida, K. Kihou, C. H. Lee, H. Kito, N. Takeshita and I. Hase, *Supercon. Sci. Technol.*, 2014, **27**, 072001
15 X. Dong and C. Fan, *Sci. Rep.*, 2015, **5**, 09326.
16 S. Hösel, *Z. Phys. Chem. (Leipzig)*, 1962, **219**, 205.
17 Y. Xu, M. Yamazaki and P. Villars, *Jpn. J. Appl. Phys.*, 2011, **50**, 11RH02.
18 T. B. Massalski and H. Okamoto, *Binary Alloy Phase Diagrams*, II vol. 1, 1990.
19 F. Merlo and M. L. Fornasini, *Mater. Res. Bull.*, 1994, **29**, 149.
20 H. Kim, D. A. Boysen, D. J. Bradwell, B. Chung, K. Jiang, A. A. Tomaszowska, K. Wang, W. Wei and D. R. Sadoway, *Electrochim. Acta*, 2012, **60**, 154.
21 S. von Náray-Szabó, *Z. Kristallogr.*, 1937, **97**, 223.
22 H. Schachner, H. Nowotny and H. Kudielka, *Monatsh. Chem. Verw. Tl.*, 1954, **85**, 1140.
23 L. M. Schoop, M. N. Ali, C. Straßer, A. Topp, A. Varykhalov, D. Marchenko, V. Duppel, S. S. P. Parkin, B. V. Lotsch and C. R. Ast, Nat. Commun., 2016, **7**, 11696
24 M. Neupane, I. Belopolski, M. M. Hosen, D. S. Sanchez, R. Sankar, M. Szlawska, Su-Yang Xu, K. Dimitri, N. Dhakal, P. Maldonado, P. M Oppeneer, D. Kaczorowski, F. Chou, M. Z. Hasan and T. Durakiewicz, Phys. Rev. B, 2016, **93**, 201104(R).
25 I. M. Chapnik, *Phys. Status. Solidi B*, 1985, **130**, K179
26 Y. R. Chung, H. H. Sung and W. H. Lee, *Phys. Rev. B*, 2004, **70**, 052511.
27 Y. Yamaguchi, S. Waki and K. Mitsugi, *J. Phys. Soc. Jpn.*, 1987, **56**, 419.
28 N. Sato, T. Kinokiri, T. Komatsubara, and H. Harima, *Phys. Rev. B*, 1999, **59**, 4714.
29 Y. Kohori, T. Kohara, N. Sato, T. Kinokiri, *Physica C*, 2003, **388-389**, 579.

30 L. L. Zhao, S. Lausberg, H. Kim, M. A. Tanatar, M. Brando, R. Prozorov and E. Morosan, *Phys. Rev. B*, 2012, **85**, 214526.
31 K. Momma and F. Izumi, *J. Appl. Crystallogr.*, 2011, **44**, 1272.
32 P. C. Canfield and Z. Fisk, *Philos. Mag. B*, 1992, **65**, 1117.
33 A. Le Bail, *Powder Diffr.*, 2005, **20**, 316.
34 H. M. Rietveld, *J. Appl. Crystallogr.*, 1969, **2**, 65.
35 J. Rodriguez-Carvajal, *Physica B*, 1993, **193**, 55.
36 P. Giannozzi, S. Baroni, N. Bonini, M. Calandra, R. Car, C. Cavazzoni, D. Ceresoli, G. L. Chiarotti, M. Cococcioni, I. Dabo, A. Dal Corso, S. de Gironcoli, St. Fabris, G. Fratesi, R. Gebauer, U. Gerstmann, C. Gougoussis, A. Kokalj, M. Lazzeri, L. Martin-Samos, N. Marzari, F. Mauri, R. Mazzarello, S. Paolini, A. Pasquarello, L. Paulatto, C. Sbraccia, S. Scandolo, G. Sclauzero, A. P. Seitsonen, A. Smogunov, P. Umari, and R. M. Wentzcovitch, *J. Phys. Condens. Matter*, 2009, **21**, 395502.
37 THEOS pseudopotential database, http://theossrv1.epfl.ch/Main/Pseudopotentials
38 J. P. Perdew, K. Burke and M. Ernzerhof, *Phys. Rev. Lett.,* 1996, **77**, 3865.
39 T. Klimczuk and R. J. Cava, *Phys. Rev. B*, 2004, **70**, 212514.
40 E. A. Lynton, *Superconductivity*, Chapman and Hall, London, 1971.
41 C. Kittel, *Introduction to Solid State Physics, 8th ed.*, John Wiley & Sons, Hoboken, 2005.
42 W. L. McMillan, *Phys. Rev.*, 1968, **167**, 331
43 A. Kokalj, *J. Mol. Graph. Model*, 1999, **17**, 176. Code available from http://www.xcrysden.org/

# SUPPLEMENTARY INFORMATION

# Superconductivity in CaBi$_2$

M.J. Winiarski[1,*], B. Wiendlocha[2], S. Gołąb[2], S. K. Kushwaha[3], P. Wisniewski[4], D. Kaczorowski[4], J. D. Thompson[5], R. J. Cava[3], T. Klimczuk[1,†]

[1] *Faculty of Applied Physics and Mathematics, Gdansk University of Technology, Narutowicza 11/12, 80-233 Gdansk, Poland*

[2] *Faculty of Physics and Applied Computer Science, AGH University of Science and Technology, Aleja Mickiewicza 30, 30-059 Krakow, Poland*

[3] *Department of Chemistry, Princeton University, Princeton NJ 08544, USA*

[4] *Institute for Low Temperature and Structure Research, Polish Academy of Sciences, PNr 1410, 50-950 Wrocław, Poland*

[5]*Los Alamos National Laboratory, Los Alamos, New Mexico 87545, USA*

\* mwiniarski@mif.pg.gda.pl

† tomasz.klimczuk@pg.gda.pl

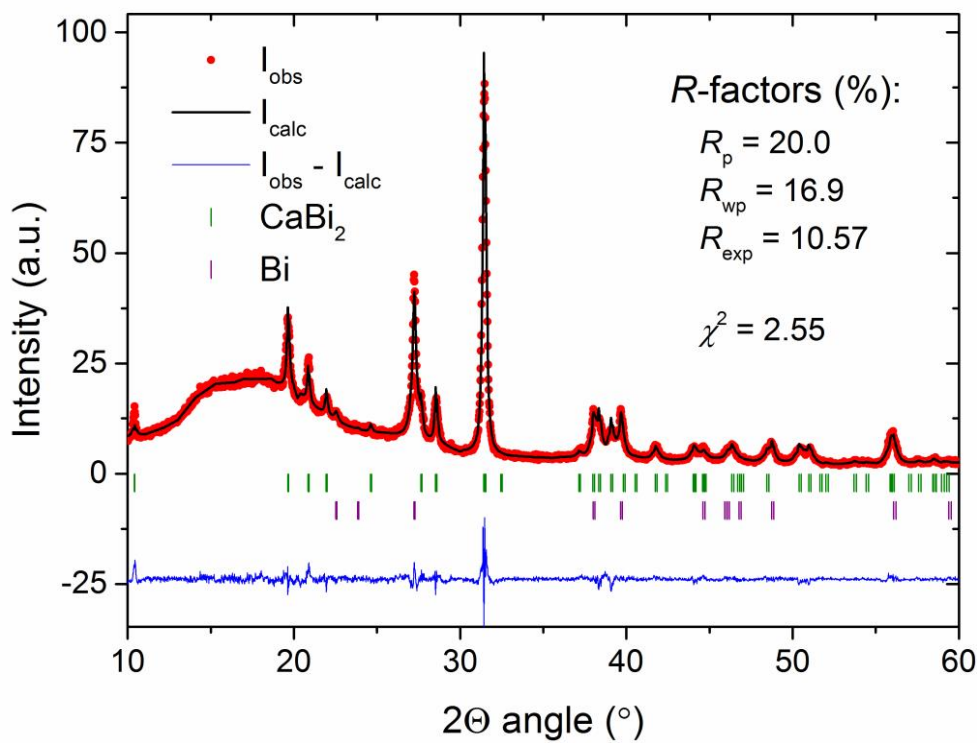

**Fig. S1.** Rietveld (CaBi$_2$ phase) and LeBail (Bi phase) fit to the diffraction pattern of ground CaBi$_2$ single crystals. Red points – observed intensities ($I_{obs}$), black solid line – calculated ($I_{calc}$), blue line – $I_{obs}$-$I_{calc}$, green and purple tics indicate the expected positions of Bragg reflections for CaBi$_2$ and Bi, respectively.

**Table S1.** Unit cell parameters and atomic positions obtained from a Rietveld fit (shown in Fig. S1). Values given in italics are repeated after Table 1 for comparison.

| Unit cell parameters (Å): | | | |
|---|---|---|---|
| Rietveld refinement | 4.6970(5) | 17.069(2) | 4.6127(4) |
| *LeBail fit (see Tab. 1)* | *4.696(1)* | *17.081(2)* | *4.611(1)* |
| *Calculated (see Tab. 1)* | *4.782* | *17.169* | *4.606* |
| Atomic positions: | | | |
|  | $x$ | $y$ | $z$ |
| Ca | 0 | 0.0974(12) | ¼ |
| Bi(1) | 0 | 0.4342(2) | ¼ |
| Bi(2) | 0 | 0.7449(3) | ¼ |
| $R$-factors (%): | | | |
| $R_p$ | $R_{wp}$ | $R_{exp}$ | $\chi^2$ |
| 20.0 | 16.9 | 10.57 | 2.55 |

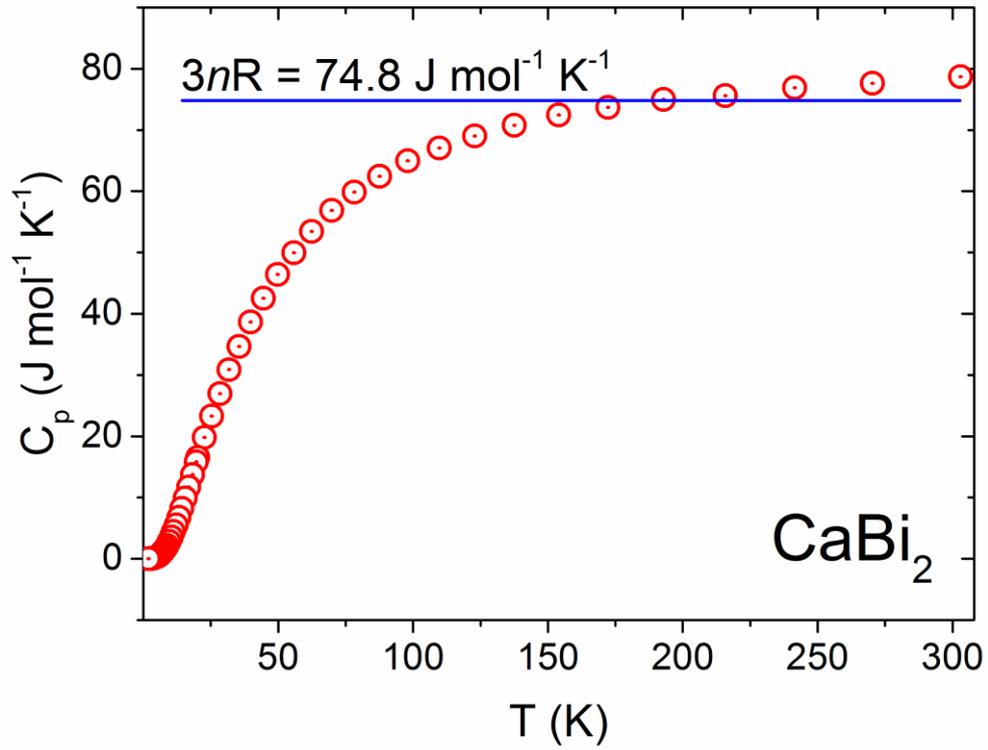

**Fig. S2.** Specific heat of CaBi$_2$ from 2 to 300 K. The solid blue line shows the value of specific heat calculated from the Dulong Petit Law assuming 3 atoms per formula unit.

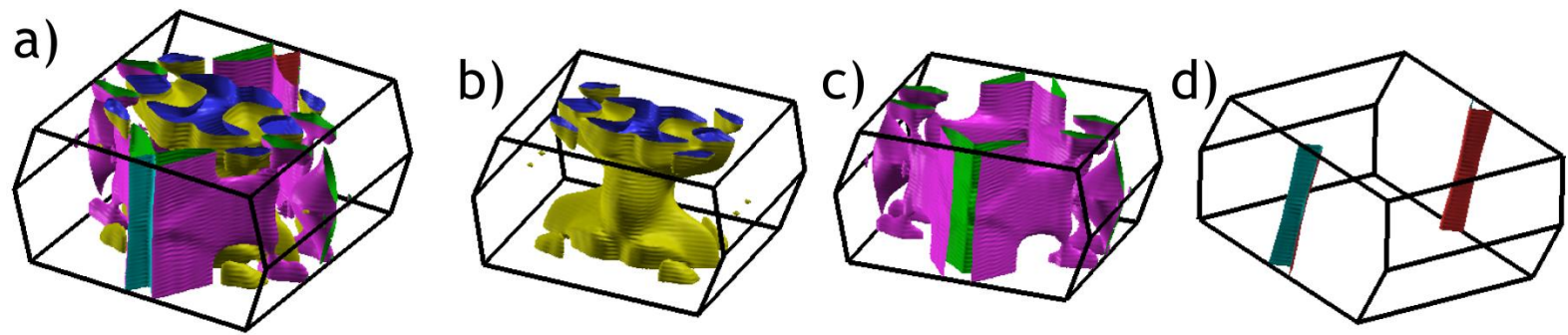
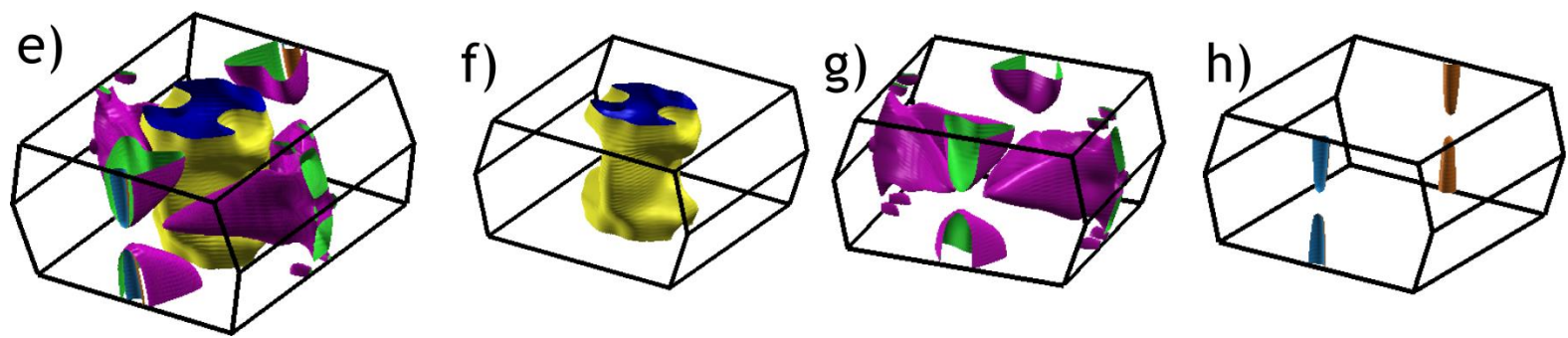

**Fig. S3.** The Fermi surface (FS) of CaBi$_2$: (a) scalar-relativistic case, (e) relativistic case; panels (b-d) and (f-h) show each of the three FS sheets in scalar-relativistic and relativistic case, respectively. Pictures prepared using XCrysDen [1].

---

[1] A. Kokalj, *J. Mol. Graph. Model*, 1999, **17**, 176. Code available from http://www.xcrysden.org/